\documentclass[preprint,12pt]{elsarticle}

\usepackage{amssymb}
\usepackage{amsmath}
\usepackage{amsfonts}       
\usepackage[utf8]{inputenc} 
\usepackage[T1]{fontenc}    
\usepackage{graphicx}		
\usepackage{url}            
\usepackage{booktabs}       
\usepackage{nicefrac}       
\usepackage{microtype}      
\usepackage{array}
\usepackage{xcolor}        
\usepackage{float}			
\usepackage{hyperref}       
\usepackage{multicol}
\usepackage{enumitem}
\usepackage{lscape}
\usepackage{multirow}
\usepackage{hhline}
\usepackage{tabularx}

\usepackage[T1]{fontenc}
\usepackage{lmodern}
\usepackage{microtype}

\journal{Nuclear Physics B}

\begin{document}

\begin{frontmatter}



\title{A Recommender System Based on Binary Matrix Representations for Cognitive Disorders}

\author[IDA]{Raoul H. Kutil} 
\author[PMU]{Georg Zimmermann}
\author[IDA]{Christian Borgelt\corref{cor1}}
\cortext[cor1]{Corresponding Author}
\ead{christian.borgelt@plus.ac.at}

\affiliation[IDA]{organization={Department of AIHI, University of Salzburg},
            addressline={Jakob Haringer Str. 2}, 
            city={Salzburg},
            postcode={5020}, 
            state={Salzburg},
            country={Austria}}
\affiliation[PMU]{organization={Research Programme Biomedical Data Science, Paracelsus Medical University},
            addressline={Strubergasse 21}, 
            city={Salzburg},
            postcode={5020}, 
            state={Salzburg},
            country={Austria}}

\begin{abstract}
\textbf{Background}: Diagnosing cognitive (mental health) disorders is a delicate and complex task. Identifying the next most informative symptoms to assess, in order to distinguish between possible disorders, presents an additional challenge. This process requires comprehensive knowledge of diagnostic criteria and symptom overlap across disorders, making it difficult to navigate based on symptoms alone.\\
\textbf{Objective}: This research aims to develop a recommender system for cognitive disorder diagnosis using binary matrix representations. \\
\textbf{Method}: The core algorithm utilizes a binary matrix of disorders and their symptom combinations. It filters through the rows and columns based on the patient’s current symptoms to identify potential disorders and recommend the most informative next symptoms to examine.\\ 
\textbf{Results}: A prototype of the recommender system was implemented in Python. Using synthetic test and some real-life data, the system successfully identified plausible disorders from an initial symptom set and recommended further symptoms to refine the diagnosis. It also provided additional context on the symptom-disorder relationships.\\
\textbf{Conclusion}: 
Although this is a prototype, the recommender system shows potential as a clinical support tool. A fully-developed application of this recommender system may assist mental health professionals in identifying relevant disorders more efficiently and guiding symptom-specific follow-up investigations to improve diagnostic accuracy.  
\end{abstract}
\smallskip




\end{frontmatter}

\section{Introduction}
Accurate diagnosis is essential for effective treatment, but the process is often complicated by the complexity and symptom overlap among mental disorders. Diagnostic manuals like the DSM-5 (Diagnostic and Statistical Manual of Mental Disorders, 5th Edition) and ICD-10 (International Classification of Diseases, 10th Edition) provide criteria meant to standardize diagnosis, but these criteria can be met in many different ways. The DSM is the standard classification system used by mental health professionals to diagnose and categorize mental disorders. It provides clear description, the previously mentioned criteria, and codes to ensure consistency and reliability in diagnosis across clinical settings. The ICD (International Classification of Diseases, 10th Edition) includes much more, but does not specifically focus on mental health issues. Each disorder includes a wide range of possible criteria-satisfying symptom combinations that fulfil the minimum requirements for diagnosis, though additional symptoms may also be present. For example, the first criterion for Major Depressive Disorder (MDD) includes nine items (consisting of over 20 symptoms), of which any five must be present, with at least one being either the symptoms depressed mood or loss of interest. \\
This results in over seven million possible valid symptom combinations. Each symptom combination is a valid configuration for satisfying this criterion of the diagnosis. A symptom combination satisfying all diagnostic criteria of a disorder will be referred to as symptom profile, or simply profile. Such variability makes diagnosis especially difficult when symptoms overlap across disorders, as seen in the shared symptoms of MDD, anxiety disorders, bipolar disorder, and certain personality disorders. This overlap complicates differential diagnosis, the process of distinguishing one disorder from others with similar symptoms. Despite these challenges, accurate diagnosis is crucial, as misdiagnosis can lead to ineffective treatment, prolonged suffering, and increased healthcare costs. While updates to diagnostic criteria, based on clinical research and practitioner feedback, have improved specificity for some conditions, the knowledge remains primarily narrative and not machine-readable. Ontologies like the Human Disease Ontology and the ICD ontology aim to structure this information, but they either reflect the authors’ interpretations or fail to represent critical relationships between symptoms and diagnostic criteria. As a result, these tools fall short of supporting scalable, computational analysis. The lack of a comprehensive, machine-readable diagnostic framework limits the ability to analyze and distinguish disorders effectively. Clinicians must make decisions under time constraints and cognitive limitations, often relying on heuristics like availability, representativeness, and confirmation bias. While these mental short cuts help manage complexity, they can introduce error, particularly when atypical symptom patterns are involved. 
Enhancing diagnostic accuracy and minimizing bias requires a balance between intuitive judgment and systematic analysis, supported by improved tools for representing and navigating diagnostic knowledge. Systems like Phen2Gene\citep{Phen2gene2019}, DXplain\citep{DXplain2019}, and Infermedica\citep{Infer2013a}\citep{Infer2013b} exemplify diverse clinical recommendation approaches, each designed to achieve specific objectives. While Phen2Gene prioritizes genes from phenotype profiles, DXplain offers knowledge-based differential diagnoses, and Infermedica dynamically refines disease rankings using probabilistic and machine-learning methods. Together, they show how recommendation strategies can be successfully applied across diverse clinical tasks.\\

Strasser-Kirchweger et al. (2025)\citep{consenscompute} introduced a novel framework that utilizes a binary system to represent the relationship between symptoms and diagnostic criteria. This representation enabled the development of a measure for assessing the similarity between cognitive disorders as described in the DSM-5\citep{dsm5}, and for exploring the potential of a delineation method between disorders. Initial experiments with the proposed measure showed alignment with expert assessments of the relative similarity among various disorders. Building on this work, Kutil et al. (2025)\citep{gens} addressed the challenges of generating symptom combinations and computing similarity measures for highly complex disorders. These issues were effectively tackled using a generator-based representation of disorders, which was originally designed to bridge the narrative descriptions with their binary matrix counterparts. 
This study proposes a prototype for a recommender system application designed to support medical professionals in navigating the diagnostic framework of the DSM-5. It implements abductive reasoning in a vectorized framework, drawing inspiration from inference-based diagnostic models proposed by Reggia et al. (1990)\citep{ReggiaPeng1990}. Abductive reasoning serves as the computational basis for deriving plausible diagnostic explanations from observed symptom patterns, consistent with its formalization as a core inference process (Selman et al., 1989)\citep{SelmanLevesque1989} and its distinction from deduction (Shanahan, 1989)\citep{Shanahan1989}. The goal is to enhance the reliability and efficiency of the diagnostic process by suggesting the most informative symptoms to evaluate next. The core idea behind the algorithm is that a clinician needs only to input a set of observed symptoms to receive suggestions for likely disorders and the next most informative symptoms to investigate, eliminating the need for additional manual research and any human error that could occur thereby.
\medskip

\subsection{Binary Matrix Representation}
A machine-actionable representation is a translation of narrative text into a format computers can efficiently process. While computers can store and read raw text from sources like the DSM-5, extracting meaningful insights is challenging due to inconsistencies in symptom naming and ambiguities in conjunctions ("and"/"or") from the diagnostic criteria. Natural language processing (NLP) helps interpret these complexities, but a direct translation into a binary system simplifies analysis. A binary matrix is such a machine-actionable representation. In this approach, symptoms are represented as binary values ($1$ for presence, $0$ for absence). The collection of symptoms as a list or as a binary vector necessary for a valid diagnosis is called a symptom profile. Two methods are considered: one that focuses solely on symptom presence/absence and another that incorporates logical rules for a more detailed representation \citep{gens}.\\

\noindent The "Maximum Profile" method (MP) represents each disorder as a single binary vector, marking all relevant symptoms with $1$s and absent ones with $0$s. Vertically stacking these vectors forms a binary matrix with disorders as rows and symptoms as columns. This profile is called maximum profile because it contains all symptoms that a relevant in any real possible manifestation of that disorder also represented by the profiles in the next method. The "All Profiles" method (AP) accounts for all valid symptom combinations, forming a separate matrix per disorder (Table \ref{tab:allprofiles}). Columns represent symptoms, while rows list the symptom profiles \citep{gens}.\\

\renewcommand{\arraystretch}{1.3}
\begin{table}[h]
\centering
\small
\begin{tabular}{|>{\bfseries}c||c|c|c|c|c|c|c|c|c|c|}
\hline
\textbf{} & \textbf{S1} & \textbf{S2} & \textbf{S3} & \textbf{S4} & \textbf{S5} & \textbf{S6} & \textbf{S7} & \textbf{S8} \\
\hline\hline
\textbf{Disorder 1}  & 0 & 0 & 1 & 1 & 1 & 1 & 0 & 1\\
\hline
\textbf{Disorder 1}  & 0 & 0 & 1 & 1 & 1 & 1 & 1 & 0\\
\hline
\textbf{Disorder 1}  & 0 & 0 & 1 & 1 & 1 & 1 & 1 & 1\\
\hline
\textbf{Disorder 2}  & 1 & 0 & 1 & 1 & 1 & 0 & 0 & 0\\
\hline
\textbf{Disorder 2}  & 0 & 1 & 1 & 1 & 1 & 0 & 0 & 0\\
\hline
\end{tabular}
\vspace{5pt}
\caption{Example of the "All Profiles" representation}
\label{tab:allprofiles}
\end{table}

\noindent Both binary matrix representations can be employed to store the disorder data locally, but the MP representation can primarily be used for initial filtering within the recommender system. The AP representation encodes all diagnostic pathways as binary symptom combinations, which the recommender system searches to identify relevant disorders and guide symptom-based differentiation.
\medskip

\subsection{Profile Generators}
Representing all symptom profiles (AP) of a disorder in a machine-actionable format becomes infeasible for complex disorders due to the rapid growth in the number of possible combinations. This explosion results from how diagnostic criteria are structured in the DSM-5\cite{dsm5} each criterion can produce multiple valid symptom combinations (e.g., "at least two of the following"), and the complete set of profiles is formed by taking the Cartesian product of these combinations across all criteria. Additional complexity arises from overlapping symptoms that count as a single criterion item, and from opposing symptoms grouped together for simplicity (e.g., insomnia and hypersomnia under "sleep problems"). As a result, some disorders may yield millions of valid profiles, making manual creation impractical due to both scale and potential for human error. To address this, a system of "profile generators" was developed, structured representations of diagnostic criteria that serve as inputs to algorithmic rules. These generators allow for the automated construction of all valid symptom profiles, forming the basis for a disorder's binary matrix representation. Each generator is a structured list that encodes the logic of DSM-5 criteria, and all of its disorders can be expressed using only these five generator types, enabling consistent and scalable transformation into a machine-readable format \citep{gens}.\\

\noindent
\textbf{Generator 0} - Identity Generator: $\quad$ [\{a,b,c,d\}]\\
Simply returns the input symptom set unchanged and is used when all symptoms are always present in every profile. It represents a special case of Generator 1, where the subset size $k$ is equal to the size of the original set $n$, and is generally redundant in practice.\\

\noindent
\textbf{Generator 1} - Subset Generator: $\quad$ [\{a,b,c,d\}, k]\\
G1 produces all subsets of a given symptom set whose size is greater than or equal to $k$. It generalizes Generator 0 and defines the basic combinatorial mechanism for generating profiles.\\

\noindent
\textbf{Generator 2} - Multi-Set Generator: $\quad$ [\{a,b\},\{c,d\},\{e,f\}, k]\\
G2 extends Generator 1 to multiple symptom sets ($S_1, \ldots, S_m$). It generates all subsets with size $\geq k$, but counts only one symptom per originating set, ensuring diversity across sets rather than within them.\\

\noindent
\textbf{Generator 3} - Disjoint Combination Generator: $\quad$ [[\{a\},\{b\}],[\{c\},\{d\}]]\\
G3 forms disjoint combinations by pairing sets from two lists $L_1$ and $L_2$ via a Cartesian product. This generator is rarely required but may be necessary for certain disorders (e.g. "Obstructive Sleep Apnea Hypopnea" from DSM-5). Its use significantly increases algorithmic complexity.\\

\noindent
\textbf{Generator 4} - Ex. Multi-Set Generator: $\quad$ [[\{a,b\},\{c,d\}],[\{e,f\},\{g\}], (r,s,t)]\\
G4 extends Generator 2 to handle disorders with subconditions. It uses a tuple $(r, s, t)$ to specify minimum symptom requirements within each list ($L_1, L_2$) and across both lists combined. This allows modeling hierarchical or compound diagnostic criteria.
\smallskip

\subsubsection{Conditional Generators}
The generator mapping (to a binary matrix) enables a straight input of two generators to calculate their MPCS, but the size of those matrices (complex disorders = many profiles) is responsible for the run time of the algorithm. A major portion of the run time is used for the matrix-matrix multiplication, which can result in a matrix that is to big to handle for common computers. This is already dealt with (in the source code of the MPCS) by switching to a matrix-vector multiplication for each row, which increases the run time but evades the problem with very large matrices. 
Although several methods can improve or optimize the base algorithm \citep{gens}, one of the most effective is to reduce the number of profiles that need to be generated in the first place. When only the most similar profiles between two disorders are of interest, their generators can be simplified by maximizing symptom overlap and minimizing symptom differences, thereby lowering the overall generation load. The following example, involving two disorders with symptoms labeled a to g, demonstrates this simplification process at a basic level.\\

\noindent
Disorder 1 = [\{a,b,c,d\}, 3]\\
Disorder 2 = [\{c,d,e,f,g\}, 4]
\begin{flushleft}
\begin{tabular}{@{}l l@{}}
Their overlap has to be \textbf{maximized}: & Disorder 1 = [\{c,d\}], [\{a,b\}, 1]\\
       & Disorder 2 = [\{c,d\}], [\{e,f,g\}, 2] \\ \\
Their difference has to be \textbf{minimized}: & Disorder 1 = [\{c,d\}], [\{a\}]\\ 
       & Disorder 2 = [\{c,d\}], [\{e,f\}] 
\end{tabular}
\end{flushleft}
This procedure, and in particular the maximization component, can function as a preliminary filtering step that restricts generation to profiles relevant to the recommender’s base algorithm, discussed in Section 2.4.
\medskip

\section{Recommender System}
A recommender system is a type of software tool designed to suggest items, content, or actions to users based on their preferences, behaviour, or other input data. Its primary aim is to assist users in discovering relevant information without the need for manual search \cite{reco1}\cite{reco2}. In this context, the recommender system suggests potential cognitive disorders and the most informative next symptoms to evaluate, based on the symptoms already provided. The goal is to support mental health professionals (and by extension, the patient) in reaching a diagnosis more efficiently. The system relies on a binary matrix representation of disorders to filter all symptom profiles, identify disorders consistent with the input symptoms, and recommend additional symptoms to further refine the differential diagnosis. The algorithm's objective is to reduce the remaining set of possible disorders to a level at which a definitive diagnosis can be made.\\

\subsection{Concept}
The algorithm operates on a binary matrix in which each row represents a symptom profile associated with a specific disorder, and each column (aside from the first) indicates the presence (1) or absence (0) of a particular symptom. The first column contains the disorder label for each profile. Note that the algorithm is agnostic to how this matrix was generated or how large it is, making it modular and suitable for various preprocessing pipelines.\\
The procedure starts by filtering the matrix so that only profiles remain which include all input symptoms (= currently observed in a patient) specified as present and exclude those specified as absent. These input symptom columns are then removed from the matrix, as the goal is to focus on what other symptoms might help differentiate between possible diagnoses. The remaining data is grouped by disorder, and for each disorder, the profiles are aggregated by computing the mean of each symptom column. This mean represents the relative frequency of each symptom within that disorder. For example, a value of 1 means the symptom is always present among the remaining profiles of that disorder, while 0 means it never appears. A value between 0 and 1 can be interpreted as the probability that the symptom is present among the remaining relevant profiles. The key idea is to use these frequencies to identify informative symptoms, symptoms that consistently appear in one disorder but never in another. Although symptoms with intermediate values are also informative, examining their role and usefulness within the algorithm is not covered in this study and remains an avenue for future investigation. The algorithm constructs two sets: one containing symptoms that are always present in at least one disorder (frequency = 1), and another for symptoms that are always absent in at least one disorder (frequency = 0). Each set individually contains valuable symptom information that enables one-sided tests for including or excluding a disorder based on the observed symptom state. However, their intersection yields an even more informative symptom set:
symptoms that are required for at least one disorder and excluded from another, and are therefore particularly useful for distinguishing between them. Additional diagnostic insight can be provided by tracing each informative symptom back to the disorders it distinguishes. This mapping clarifies the discriminative role of each symptom and strengthens the interpretability of the algorithm's output.\\

\subsection{Definitions}
\vspace{-15pt}
{\small
\begin{align*}
s_i \quad &\ldots \quad \textrm{Symptom}\\
p_i=\{s_1,\ldots,s_n\} \quad &\ldots \quad \textrm{Profile (Set of Symptoms)}\\
D_i=\{p_1,\ldots, p_m\} \quad &\ldots \quad \textrm{Disorder (Set of Profiles)}\\
E_i=\{D_1,\ldots, D_l\}\quad &\ldots \quad \textrm{Set of Disorders}\\
Z = \{z_1,\ldots,z_r\} \quad &\ldots \quad \textrm{Given Symptoms (Set of Symptoms)}
\end{align*}}\\

\noindent The combined symptom space for each disorder $D_i$ for the algorithm is defined by
\begin{align}
B = \bigcup_{i} \{ p_i \mid p_i \in D_i \}
\end{align}
The necessary profiles for each disorder based on the given symptoms are calculated with
\begin{align}
D^{'}_i = \{ p\backslash Z \mid p\in D_i, p\cap Z = Z \}
\end{align}

\noindent The reduced disorders are transformed into binary matrices with the combined symptom space $B$ in mind:
{\small
\begin{align*}
i,j,k\in \mathbb{N}\quad &\ldots \quad \textrm{Index of Profile, Symptom, Disorder}\\
l \in \mathbb{N}\quad &\ldots \quad \textrm{Number of Disorders}\\
a_{ij}\in \{0,1\} \quad &\ldots \quad \textrm{Symptom Identifier (row i, column j)}\\
M_{D^{'}} = (a_{ij})_B \in \{0,1\}^{n \times m} \quad &\ldots \quad \textrm{Binary Matrix Transform of Disorder }D\\
n:=|M_{D^{'}}|=\textrm{rows}(M_{D^{'}}) \quad &\ldots \quad \textrm{Number of Rows}\\
m:=|B|=\textrm{cols}(M_{D^{'}}) \quad &\ldots \quad \textrm{Number of Columns}
\end{align*}}

\noindent Each disorder $D_k$ transform with a column aggregation to
\begin{align}
C_j = \frac{1}{n} \sum^{n}_{i=0} a_{ij} \quad \textrm{with } C_j \in [0,1]
\end{align}
\begin{align}
\overline{M_{D^{'}}} := (C_1,\ldots, C_m)
\end{align}

\noindent For each index $k$ of disorder $D_k$ in $E$, we compute the indices of the columns for the $1$- and $0$-sets:
\begin{align}
S^{k}_1 &= \{i \mid i \in \{1,\ldots,m\}, \exists C_i \in \overline{M_{D^{'}_k}}:C_i = 1 \}\\
S^{k}_0 &= \{i \mid i \in \{1,\ldots,m\}, \exists C_i \in \overline{M_{D^{'}_k}}:C_i = 0 \}\\
S_{\textrm{inter}} &=\bigcup S^{k}_1 \cap \bigcup S^{k}_0
\end{align}
The final set $S_{\textrm{inter}}$ contains all indices of columns (= symptoms) for which a hard distinction can be made. This hard distinction is the definite inclusion or exclusion for specific disorders based on the symptoms represented by an index of this set.

\subsection{Base Algorithm}
\noindent \textbf{Step 1}\\
First, any row that does not contain the required $1$s for input symptoms specified as present, or the required $0$s for input symptoms specified as absent, is removed from the matrix. Next, the columns representing the input symptoms themselves are removed. The remaining rows are then grouped by disorder name (assumed to be in the first column), and symptom combinations belonging to the same disorder are aggregated. The first output of the algorithm is a list of disorder names corresponding to these aggregated groups.\\

\noindent \textbf{Step 2}\\
Aggregation is performed by computing the mean value of each column across all profiles within a disorder group. This yields the relative frequency of each symptom among the remaining profiles for that disorder. These relative frequencies are used to categorize symptoms into three sets:
\begin{itemize}
\item $S_1$: symptoms that appear with a value of 1 (aggregation) in at least one disorder
\item $S_0$: symptoms that appear with a value of 0 (aggregation) in at least one disorder
\item $S_I=S_1 \cap S_0$: the intersection of the two sets, representing symptoms that are always required in one disorder and never in another
\end{itemize}
The second output of the algorithm is the set $S_I$, which contains the most informative symptoms for distinguishing between the remaining candidate disorders. The symptoms sets $S_1$ and $S_0$ serve as secondary outputs of this step and become relevant if $S_I$ is empty.\\

\noindent \textbf{Step 3}\\
The final step involves backtracking from the results of Step 2 to identify which specific disorders each symptom in the intersection set can distinguish. This step serves as supplementary information, providing additional context for the distinguishing symptoms identified earlier.\\
\medskip

The following synthetic example illustrates how the algorithm works. The disorders (D1, D2, D3 and D4) and their associated symptom combinations are fictional and do not reflect actual symptom configurations from the DSM-5. Each disorder is assigned a different number of profiles, which may change as filtering steps are applied during the algorithm's execution.\\
At first,  all disorders are combined into a single matrix (Table \ref{tab:firstbin}), which results in a total of 10 symptoms, labeled $S1$ through $S10$. Initially, Disorder 1 has five profiles, Disorder 2 has three, Disorder 3 has two, and Disorder 4 has one. At the start of the procedure, symptoms $S5$ through $S8$ are provided to the algorithm as input, indicating the symptoms already known to be present.
\renewcommand{\arraystretch}{1.3}
\begin{table}[h]
\centering
\footnotesize
\begin{tabular}{|>{\bfseries}c||c|c|c|c|c|c|c|c|c|c|}
\hline
\textbf{Disorder} & \textbf{S1} & \textbf{S2} & \textbf{S3} & \textbf{S4} & \textcolor{blue}{\textbf{S5}} & \textcolor{blue}{\textbf{S6}} & \textcolor{blue}{\textbf{S7}} & \textcolor{blue}{\textbf{S8}} & \textbf{S9} & \textbf{S10} \\
\hline\hline
\textbf{D1}  & 1 & 1 & 1 & 0 & 1 & 1 & 1 & 1 & 0 & 1\\
\hline
\textbf{D1}  & 1 & 1 & 0 & 0 & 1 & 1 & 1 & 1 & 1 & 0\\
\hline
\textbf{D1}  & 1 & 0 & 0 & 0 & 1 & 1 & 1 & 1 & 1 & 0\\
\hline
\textbf{D1}  & 0 & 0 & 0 & 0 & 1 & 1 & 1 & 0 & 0 & 0\\
\hline
\textbf{D1}  & 0 & 0 & 0 & 0 & 1 & 1 & 0 & 1 & 0 & 0\\
\hline
\textbf{D2}  & 0 & 0 & 1 & 1 & 1 & 1 & 1 & 1 & 0 & 1\\
\hline
\textbf{D2}  & 0 & 0 & 0 & 0 & 0 & 1 & 1 & 1 & 0 & 0\\
\hline
\textbf{D2}  & 0 & 0 & 1 & 0 & 1 & 1 & 1 & 1 & 0 & 0\\
\hline
\textbf{D3}  & 1 & 0 & 0 & 1 & 1 & 1 & 1 & 1 & 1 & 1\\
\hline
\textbf{D3}  & 0 & 0 & 0 & 0 & 1 & 1 & 1 & 0 & 0 & 0\\
\hline
\textbf{D4}  & 1 & 0 & 0 & 0 & 1 & 0 & 0 & 1 & 0 & 1\\
\hline
\end{tabular}
\vspace{5pt}
\caption{Initial binary matrix with all profiles}
\label{tab:firstbin}
\end{table}
After the first filtering step, Disorder 1 retains three profiles, Disorder 2 has two, Disorder 3 has one, and Disorder 4 is fully excluded. In the second step, grouping and aggregating (mean) the remaining profiles by disorder produces the data in the following table (Table \ref{tab:finalbin}).\\
\begin{table}[h]
\centering
\small
\begin{tabular}{|>{\bfseries}c||c|c|c|c|c|c|c|c|c|c|}
\hline
\textbf{Disorder} & \textbf{S1} & \textbf{S2} & \textbf{S3} & \textbf{S4} & \textbf{S9} & \textbf{S10} \\
\hline\hline
\textbf{D1}  & 1.0 & 0.6 & 0.3 & 0.0 & 0.6 & 0.3\\
\hline
\textbf{D2}  & 0.0 & 0.0 & 1.0 & 0.5 & 0.0 & 0.5\\
\hline
\textbf{D3}  & 1.0 & 0.0 & 0.0 & 1.0 & 1.0 & 1.0\\
\hline
\end{tabular}
\vspace{5pt}
\caption{Processed binary matrix with aggregated group profiles}
\label{tab:finalbin}
\end{table}

\noindent It is evident that in columns S1, S3, S4, and S9, the aggregated values vary across disorders, with some disorders showing a 0 and others a 1. Therefore, we define the sets:
\begin{itemize}
\item $S_1=\{S1,S3,S4,S9,S10\}$: symptoms present (value = 1) in at least one disorder
\item $S_0=\{S1,S2,S3,S4,S9\}$: symptoms absent (value = 0) in at least one disorder
\item $S_I=S_1\cap S_0=\{S1,S3,S4,S9\}$: symptoms that occur in some disorders but not others
\end{itemize}
The intersection set $S_I$ identifies the most informative symptoms for further narrowing the list of candidate disorders. By backtracking these symptoms in the filtered binary matrix, the algorithm can determine which specific disorders each symptom helps to distinguish.\\

\noindent \textbf{Algorithm Output}:
{\small
\begin{itemize}[left=0pt]
\setlength\itemsep{0pt}
\item Possible Disorders: Disorder 1, Disorder 2, Disorder 3
\item Best Symptoms: Symptom 1, Symptom 2, Symptom 3, Symptom 4
	\begin{itemize}
	\item Useful symptoms for present (1): $S_1=\{S1,S3,S4,S9,S10\}$
	\item Useful symptoms for absent (0):  $S_0=\{S1,S2,S3,S4,S9\}$
	\end{itemize}
\item Symptom - Disorder - Differ: 
	\begin{itemize}
	\setlength\itemsep{0pt}
	\item Symptom 1 corresponds to Disorder 3/1 and 3/2
	\item Symptom 3 corresponds to Disorder 2/3
	\item Symptom 4 corresponds to Disorder 3/4
	\item Symptom 9 corresponds to Disorder 3/2
	\end{itemize}
\end{itemize}}

\smallskip
\subsection{Extension}
The proposed extension to the base algorithm preserves its core computational semantics while focusing on pre-processing procedures that govern data selection, structured loading, and transformation into a standardized binary matrix representation amenable to downstream processing.
This extension leverages the generator-based framework introduced in Section 1.3 to address scalability challenges arising from disorders with highly complex profile spaces. In such cases, even after conventional symptom-based filtering, the binary matrices associated with candidate disorders may exceed feasible computational limits. To mitigate this, disorder profiles are not enumerated a priori but are instead instantiated dynamically via conditional generators, as formalized in Kutil et al. (2025)\citep{gens}.\\

\noindent These conditional generators utilize the user's input symptom set to constrain the generation process, enabling the early exclusion of infeasible or redundant profiles without explicit materialization. The simplification strategy uses the generator maximization heuristic (Section 1.3) to generate a small but representative set of profiles, thereby improving both runtime efficiency and memory usage during the generation phase.\\

\begin{figure}[H]
	\centering
	\includegraphics[width=1\textwidth]{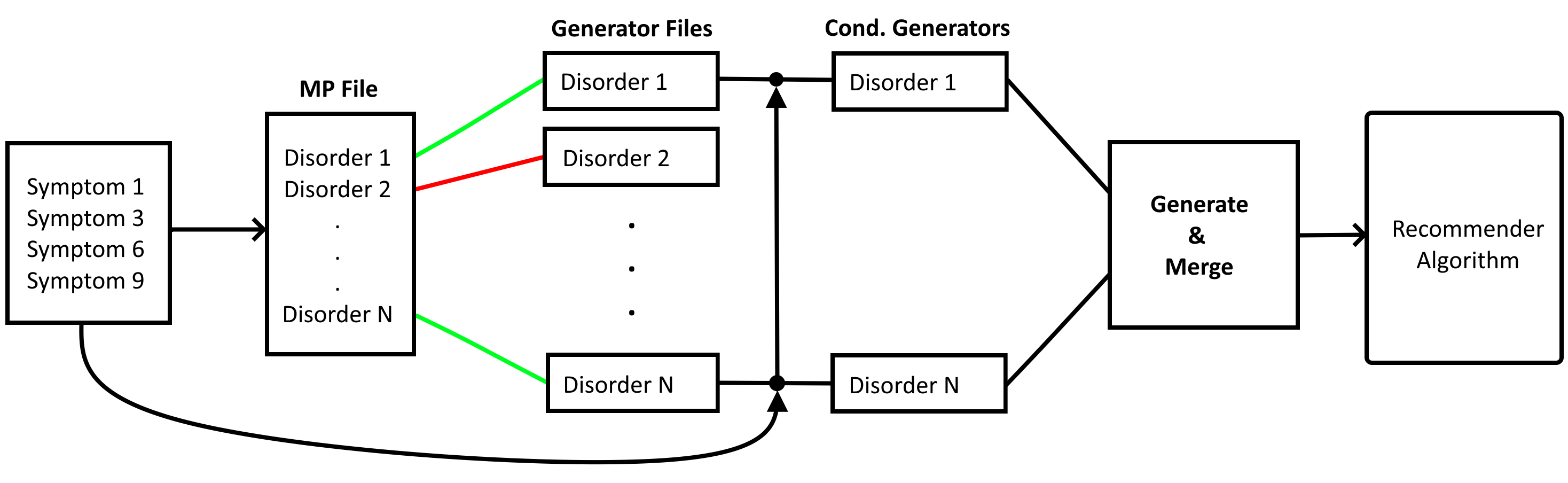}
	\caption{Pipeline of the extension with generators}
	\label{fig:4}
\end{figure} \vspace{-5pt}

The following example demonstrates how applying input-based simplification (or maximization) to generators can significantly reduce the number of profiles that need to be generated. Both illustrative disorders are modeled by a single generator ($g1, g2$) and consist of eight symptoms each, ranging from $a$ to $h$ and $d$ to $k$, with five symptoms overlapping. The two disorders differ, however, in the number of symptoms that must be present to form a valid symptom combination.\\

$
\begin{array}{l@{\;}r@{\;}llr}
g1 =& \text{[\{a, b, c, } & \text{d, e, f, g, h\}, 5]} &-& \text{ 93 profiles}\\
g2 =& \text{[\{} & \text{d, e, f, g, h, i, j, k\}, 4]} &-& \text{ 163 profiles}
\end{array}
$\\

\begin{flushleft}
\begin{tabular}{@{}l l l@{}}
Case 1:& Input symptoms $\{d, f , h\}$ & \\
       & $g1$ = [\{d, f, h\}],[\{a, b, c, e, g\}, 2] $\quad$ & $-\quad$ 26 profiles\\
       & $g2$ = [\{d, f, h\}],[\{e, g, i, j, k\}, 1] $\quad$ & $-\quad$ 31 profiles\\ \\
Case 2:& Input symptoms $\{d, f , h, g\}$ & \\
       & $g1$ = [\{d, f, g, h\}],[\{a, b, c, e\}, 1] $\quad$ & $-\quad$ 15 profiles\\
       & $g2$ = [\{d, f, g, h\}],[\{e, i, j, k\}, 0] $\quad$ & $-\quad$ 16 profiles
\end{tabular}
\end{flushleft}
\medskip
This approach, used together with the input symptoms, serves only as a pre-filtering step to reduce the size of each disorder’s matrix. Therefore, the input symptoms must be included in the disorder’s generator representation. However, if Generator 3 (see Sect. 1.2) is used, special case handling becomes necessary. When the input symptoms, or portions of them, appear in separate sets within the same list in Generator 3, the disorder can be disregarded entirely; otherwise, it continues to the next stage of processing.\\

\noindent While this generator-based approach enables efficient handling of large and structurally intricate disorders, its efficacy is contingent upon the informativeness of the input symptom set. Moreover, although the resulting matrix dimensionality is significantly reduced relative to full enumeration, profile generation remains computationally intensive in scenarios involving highly entangled or weakly constrained symptom-disorder mappings.\\
\medskip

\section{Results}

The algorithm was designed for a recommender system that represents disorders as binary matrices, where symptoms are encoded as presence ($1$) or absence ($0$) values. Each stage of the algorithm can be formulated mathematically, enabling theoretical validation of its logic and structure. The central assumption underlying the method is that, in cases without comorbidity, any two disorders must be distinguishable by at least one symptom. The implementation of the base algorithm was validated through a series of synthetic data experiments designed to emulate diverse diagnostic configurations. One such scenario, presented in Section 2.3, illustrates the general evaluation approach. Across all test cases, the algorithm successfully filtered and aggregated symptom profiles, identified plausible candidate disorders, and proposed informative distinguishing symptoms. These outcomes demonstrate the logical consistency and computational robustness of the core system architecture.\\
The extended component, which is responsible for generating simplified AP matrices from disorder generators via symptom-based maximization, is currently in the evaluation phase. Although comprehensive end-to-end validation has not yet been completed, all constituent modules, including profile generation and filtering, have been individually tested and shown to operate as intended. Ongoing experiments are focused on assessing performance under more complex, real-world diagnostic conditions.\\
Taken together, these results confirm the viability of a functional prototype for a modular recommender system. 
This prototype establishes a solid foundation for subsequent development toward real-time and large-scale clinical applications, contingent on further optimization and full integration of the extended generation module.
\smallskip

\section{Discussion}
The working prototype of the proposed recommender system performs as expected when applied to both synthetic data and limited real-world cases in controlled testing environments. Its modular design allows for efficient integration of pre-processing steps such as encoding, loading, and decoding of symptom profiles in binary matrix form. While the limited scale of available real-life data constrains broad generalizability, the results demonstrate a solid conceptual and technical foundation for a fully functional system. 
Additionally, the conditional generator extension offers a promising fallback strategy, dynamically producing profile data when input symptoms are too sparse for effective filtering or when the initial disorder set is too broad to constrain efficiently.
\medskip

\section{Conclusion}
We developed a modular recommender system for cognitive disorders that suggests likely diagnoses and follow-up symptoms to help medical professionals refine diagnostic decisions. The working prototype demonstrated correct behaviour on synthetic datasets, successfully leveraging its filtering components to generate accurate recommendations. A binary matrix representation of disorders and symptom profiles forms the basis of the system, enabling efficient data handling and interpretation.\\
In addition to the core implementation, we introduced an extension to dynamically generate profile data on the spot when input symptoms are too sparse for effective filtering or when the initial disorder set is too broad to narrow down efficiently for the algorithm. This underscores the system's adaptability, modularity and support its potential scalability.\\
The results show that a recommender algorithm based on binary symptom matrices, when fully developed, could support faster and more consistent diagnostic processes, not only for cognitive disorders, but also for any domain that can be modelled in a similar form. Future work will focus on transforming the prototype into a real-world application using clinically validated disorder profiles and advancing the extension incorporating it into the system.
\medskip


\section*{Code Availability}
The code used to generate the results in this paper is available in a public GitHub repository at: \\
\url{https://github.com/raoul-k/AIDA-Path/tree/main/Python/Recommender}

\section*{Declarations}
\subsection*{Abbreviations}
\noindent
\textbf{DSM}: Diagnostic and Statistical Manual of Mental Disorders \\
\textbf{ICD}: International Classification of Diseases \\
\textbf{NLP}: Natural Language Processing\\
\textbf{BM}: Binary Matrix \\
\textbf{AP}: All Profiles \\
\textbf{MP}: Maximum Profile \\
\textbf{MPCS}: Maximum Pairwise Cosine Similarity \\
\textbf{MDD}: Major Depressive Disorder \\
\textbf{PDD}: Persistent Depressive Disorder \\
\textbf{CPU}: Central Processing Unit\\
\textbf{GPU}: Graphics Processing Unit\\
\textbf{CUDA}: Compute Unified Device Architecture\\

\subsection*{Acknowledgements}
\noindent
We are grateful to Florian Hutzler (Department of Psychology, Centre for Cognitive Neuroscience, University of Salzburg) for his insightful feedback and valuable discussions throughout the development of this work. We would also like to thank our colleagues at the IDA Lab Salzburg for their support, valuable discussions, and minor contributions to this work.\\

\subsection*{Funding}
\noindent
All authors gratefully acknowledge the support of the InnovationExpress 2021 project {AIDA-PATH} (20102-F2101312-FPR). GZ gratefully acknowledges the support of the WISS 2025 projects 'IDA-Lab Salzburg' (20204-WISS/225/197-2019 and 20102-F1901166-KZP) and 'EXDIGIT' (Excellence in Digital Sciences and Interdisciplinary Technologies) (20204-WISS/263/6-6022)\\

\subsection*{Declaration of generative AI and AI-assisted technologies in the writing process}
\noindent
During the preparation of this work the author(s) used ChatGPT to improve readability and language. After using this tool, the author(s) reviewed and edited the content as needed and take(s) full responsibility for the content of the published article.\\


\smallskip

\bibliographystyle{unsrt}
\bibliography{recommender_bibliography.bib}

\begin{thebibliography}{10}

\bibitem{Phen2gene2019}
Fang L Chen Y Peng J Liu C Wu C Sarmady M Botas P Isla J Lyon GJ Weng C Wang~K.
  Zhao~M, Havrilla~JM.
\newblock Phen2gene: rapid phenotype-driven gene prioritization for rare
  diseases.
\newblock {\em NAR Genomics and Bioinformatics}, 2(2), 2020.

\bibitem{DXplain2019}
Hupp JA Hoffer~EP. Barnett~GO, Cimino~JJ.
\newblock Dxplain. an evolving diagnostic decision-support system.
\newblock {\em Journal of the American Medical Association}, 258(1):67--74,
  1987.

\bibitem{Infer2013a}
Hołownia~K. Zagorecki~A, Orzechowski~P.
\newblock A system for automated general medical diagnosis using bayesian
  networks.
\newblock {\em Studies in Health Technology and Informatics},
  192(5):461–--465, 2013.

\bibitem{Infer2013b}
Hołownia~K. Zagorecki~A, Orzechowski~P.
\newblock Online diagnostic system based on bayesian networks.
\newblock In {\em Conference on Artificial Intelligence in Medicine in Europe},
  2013.

\bibitem{consenscompute}
B.~Strasser-Kirchweger, R.~H. Kutil, G.~Zimmermann, C.~Borgelt, W.~Trutschnig,
  and F.~Hutzler.
\newblock Machine-actionable criteria chart the symptom space of mental
  disorders.
\newblock \url{https://doi.org/10.1101/2025.09.12.25335630}, 2025.
\newblock medRxiv preprint.

\bibitem{dsm5}
{American Psychiatric Association}.
\newblock {\em Diagnostic and Statistical Manual of Mental Disorders}.
\newblock American Psychiatric Publishing, Arlington, VA, 5th, text revision
  edition, 2022.

\bibitem{gens}
R.~H. Kutil, G.~Zimmermann, B.~Strasser-Kirchweger, and C.~Borgelt.
\newblock Profile generators: A link between the narrative and the binary
  matrix representation.
\newblock \url{https://doi.org/10.1101/2023.05.12.123456}, 2025.
\newblock arXiv preprint.

\bibitem{ReggiaPeng1990}
James~A. Reggia and Yun Peng.
\newblock {\em Abductive Inference Models for Diagnostic Problem-Solving}.
\newblock Springer, New York, NY, USA, 1990.

\bibitem{SelmanLevesque1989}
Bart Selman and Hector~J. Levesque.
\newblock Abductive and default reasoning: A computational core.
\newblock In {\em Proceedings of the 8th International Joint Conference on
  Artificial Intelligence (IJCAI-89)}, pages 343--348, Detroit, Michigan, USA,
  1989. Morgan Kaufmann.

\bibitem{Shanahan1989}
Murray Shanahan.
\newblock Prediction is deduction but explanation is abduction.
\newblock In {\em Proceedings of the 11th International Joint Conference on
  Artificial Intelligence (IJCAI-89)}, pages 1055--1060, Detroit, Michigan,
  USA, 1989. Morgan Kaufmann.

\bibitem{reco1}
C.C. Aggarwal.
\newblock {\em Recommender systems: The textbook}.
\newblock Springer, 2016.

\bibitem{reco2}
F.~Ricci, L.~Rokach, and B.~Shapira.
\newblock {\em Introduction to recommender systems handbook}.
\newblock Springer, 2011.

\end{thebibliography}

\end{document}